\documentstyle[preprint,aps]{revtex}
\tighten
\draft
\begin{document}
\preprint{\today}
\title{A quantitative test of the mode-coupling theory of the ideal
glass transition for a binary Lennard-Jones system}
\author{Markus Nauroth and Walter Kob\cite{wkob}}
\address{Institut f\"ur Physik, Johannes Gutenberg-Universit\"at,
Staudinger Weg 7, D-55099 Mainz, Germany}
\maketitle

\begin{abstract}
Using a molecular dynamics computer simulation we determine the
temperature dependence of the partial structure factors for a binary
Lennard-Jones system. These structure factors are used as input data to
solve numerically the wave-vector dependent mode-coupling equations in
the long time limit.  Using the so determined solutions, we compare the
predictions of mode-coupling theory (MCT) with the results of a
previously done molecular dynamics computer simulation [Phys.  Rev. E
{\bf 51}, 4626 (1995), {\it ibid.} {\bf 52}, 4134 (1995)]. From this
comparison we conclude that MCT gives a fair estimate of the critical
coupling constant, a good estimate of the exponent parameter, predicts
the wave-vector dependence of the various nonergodicity parameters very
well, except for very large wave-vectors, and gives also a very good
description of the space dependence of the various critical
amplitudes.  In an attempt to correct for some of the remaining
discrepancies between the theory and the results of the simulation, we
investigate two small (ad hoc) modifications of the theory. We find
that one modification gives a worse agreement between theory and
simulation, whereas the second one leads to an improved agreement.

\end{abstract}

\narrowtext


\section{Introduction}
\label{sec1}

In the last few years strong evidence was given that for certain types
of glass formers {\it at least} the universal predictions of the
so-called mode-coupling theory (MCT) are correct in that it was shown
that, e.g., there exists a critical temperature $T_c$, that the
factorization property in the $\beta$-relaxation regime holds, or that
two distinct diverging time scales can be observed. An introduction to
the theory can be found in some recent review
articles~\cite{bibles,schilling92,cummins94} and in Ref.~\cite{yip95}
the reader will find a compendium of many investigations performed to
test the validity of the theory. The outcome of most of these tests is
that, at least for fragile glassformers, MCT is a valid theory,
although recent calculations have shown that the theory might even be
applicable to relatively strong glassformers, such as
glycerol~\cite{franosch96}.  Apart from some noticeable exceptions,
discussed below, most of the tests done to check the validity of the
theory investigated only whether the {\it universal} predictions of the
theory are correct. The reason for this is the fact that the {\it
nonuniversal} predictions of the theory, e.g. the value of $T_c$ or the
details of the wave vector dependence of the nonergodicity parameters,
can be tested only for those systems for which the temperature
dependence of the structure factor (or of the partial structure factors
in the case of a multi component system) is known with a fairly high
accuracy. Since in most cases these structure factors were not
available with the required accuracy, only the universal predictions of
the theory could be tested.  The drawback with these types of tests is
that the various (nonuniversal) parameters occurring in the theory,
such as the exponent parameter $\lambda$, the critical temperature
$T_c$, or the nonergodicity parameter $f_c(q)$, had to be considered as
fit parameters of the theory, thus making the tests less stringent.
There are two exceptions to this. The first one is a system of
colloidal particles whose glass transition was studied extensively in
light scattering experiments by Pusey, van Megen and
Underwood~\cite{colloids}.  The structure of such systems is believed
to be modelled well by a system of hard spheres. For the structure
factor of the latter very reliable analytical expressions are
available~\cite{fluid_books} and thus were used by G\"otze and
Sj\"ogren to demonstrate that in the $\beta$-relaxation regime the
dynamics of the colloidal particles could be described very well with
MCT~\cite{goetze91}. The second system is a model of soft spheres for
which Barrat and Latz showed~\cite{barrat90} that MCT gives a fair
quantitative description of quantities like the nonergodicity parameter
and the critical coupling constant, which were determined by means of
computer simulations~\cite{bernu87,roux89}. Thus these two examples
show that MCT is able to give not only a {\it qualitative} correct
description of the dynamics of supercooled liquids, but that, {\it at
least} in some cases, it also gives a {\it quantitative} correct
description.

In a recent computer simulation we studied the dynamics of a mixture of
Lennard-Jones particles~\cite{kob94,kob95a,kob95b,kob95c,kob95d}. It
was shown that at low temperatures the dynamics of this system could be
described very well by MCT. However, in these papers only the universal
properties of the theory were tested since we used the data from the
simulation to fit the occurring parameters of the theory. The goal of
the present paper is now to compare the results of the simulation with
the predictions of the theory without using any fit parameter at all.
The only input to the theory will be the partial structure factors that
were obtained from the simulation. In this way it will be possible to
make a more stringent test of the theory than it was done in
Refs.~\cite{kob94,kob95a,kob95b,kob95c,kob95d} and therefore to test
whether also for this system the theory is correct not only in a
qualitative way but also in a quantitative way.

The rest of the paper is organized as follows: In Sec.~\ref{sec2} we
collect the MCT equations needed to compute the various quantities we
investigate. In Sec.~\ref{sec3} we give the details of the model and
of the numerical calculations. In Sec.~\ref{sec4} we present our results
and summarize and discuss them in Sec.~\ref{sec5}.

\section{Mode-Coupling Theory}
\label{sec2}

In this section we summarize the equations that are necessary to
describe the dynamics of the system within the framework of MCT. Most
of these equations can also be found in
Refs.~\cite{barrat90,gotze85,gotze87,bosse87,fuchs93,fuchs_phd}.

We consider a two-component system of classical particles with particle
concentrations $x_i$ and masses $m_i$, $i=1,2$.  In the following the
dynamics of the system will be described by means of the partial
intermediate scattering functions
\begin{equation}
F_{ij}(q,t)=\langle \delta\rho_i(q,0)\delta\rho_j^{\star}(q,t)\rangle\quad ,
\label{eq0}
\end{equation}
where $\delta\rho_i(q,t)$ is the density fluctuation for wave vector
$q$ at time $t$ of species $i$.  For a binary system it is useful to
collect these functions in a $2\times 2$ matrix ${\bf F}(q,t)$ with
$[{\bf F}(q,t)]_{ij}=F_{ij}(q,t)$. The equation of motion of ${\bf F}$
is given by
\begin{equation}
\ddot{{\bf F}}(q,t)+{\mbox{\boldmath $\Omega $}}^2(q){\bf F}(q,t)+
\int_0^td\tau {\bf M}(q,t-\tau) \dot{{\bf F}}(q,\tau) = 0 \quad ,
\label{eq1}
\end{equation}
where the frequency matrix ${\mbox{\boldmath $\Omega $}}^2$ is given by
\begin{equation}
\left[{\mbox{\boldmath $\Omega $}}^2(q)\right]_{ij}=q^2k_B T 
(x_i/m_i)\sum_{k}\delta_{ik} \left[{\bf S}^{-1}(q)\right]_{kj}\quad.
\label{eq2}
\end{equation}
Here ${\bf S}(q)$ stands for the $2\times 2$ matrix consisting of the
partial structure factors $S_{ij}(q)$.

Within the mode-coupling approximation the memory term ${\bf M}$ is
given at long times by
\begin{equation}
M_{ij}({\bf q},t)=\frac{k_B T}{2n m_i x_j}\int\frac{d 
{\bf k}}{(2\pi)^3}
\sum_{\alpha\beta}\sum_{\alpha'\beta'}V_{i\alpha\beta}({\bf q},{\bf k})
V_{j\alpha'\beta'}({\bf q},{\bf q-k}) F_{\alpha\alpha'}({\bf k},t)
F_{\beta\beta'}({\bf q-k},t)\quad ,
\label{eq3}
\end{equation}
where $n$ is the particle density and the vertex 
$V_{i\alpha\beta}({\bf q},{\bf k})$ is given by
\begin{equation}
V_{i\alpha\beta}({\bf q},{\bf k})=\frac{{\bf q}\cdot {\bf
k}}{q}\delta_{i\beta} c_{i\alpha}({\bf k})+
\frac{{\bf q}\cdot ({\bf q}-{\bf k})}{q} \delta_{i\alpha} c_{i\beta}
({\bf q}-{\bf k})
\label{eq4}
\end{equation}
and the matrix of the direct correlation function is defined by
\begin{equation}
c_{ij}({\bf q})=\frac{\delta_{ij}}{x_i}-
\left[{\bf S}^{-1}({\bf q})\right]_{ij} \quad .
\label{eq5}
\end{equation}

Making use of the isotropy of the system the expression for 
$M_{ij}({\bf q},t)$ can be reduced to a two dimensional integral:
\begin{eqnarray}
M_{ij}(q,t) & = & \frac{k_B T}{32n x_j m_i\pi^2 q^3}\int_0^{\infty}
dk k \int_{|q-k|}^{q+k} dp p \sum_{\alpha\beta}\sum_{\alpha' \beta'}
F_{\alpha\alpha'}(q,t) F_{\beta\beta'}(p,t) \cdot \nonumber \\
& & \left\{ \left(q^2+k^2-p^2\right) \delta_{i\beta}c_{i\alpha}(k)+
\left(q^2+p^2-k^2\right)\delta_{i\alpha}c_{i\beta}(p)\right\} \cdot 
\nonumber \\
& & \left\{ \left(q^2+k^2-p^2\right)\delta_{j\beta'}c_{j\alpha'}(k)+
\left(q^2+p^2-k^2\right)\delta_{j\alpha'}c_{j\beta'}(p)\right\} 
\quad .
\label{eq6}
\end{eqnarray}

The memory function for the incoherent intermediate scattering
function $F_i^{(s)}$ is given by:
\begin{eqnarray}
M_{i}^{(s)}({\bf q},t) & = & \int \frac{d{\bf k}}{(2\pi)^3} \frac{1}{n} 
\left(\frac{{\bf q}\cdot {\bf k}}{q}\right) (cF)_i ({\bf k},t)
F_{i}^{(s)}({\bf q}-{\bf k},t) \nonumber \\
& = & \frac{1}{16\pi^2n q^3}\int_0^{\infty} dk \int_{|q-k|}^{q+k}
dp p \left\{q^2+k^2-p^2\right\}^2 (cF)_i(k,t) F_{i}^{(s)}(p,t)\quad ,
\label{eq7}
\end{eqnarray}
with
\begin{equation}
(cF)_i(k,t)=(c_{ii}(q))^2 F_{ii}(q,t)+2c_{ii}(q)c_{ij}(q)F_{ij}(q,t)
+(c_{ij}(q))^2F_{jj}(q,t)\quad j\neq i \quad .
\label{eq8}
\end{equation}

The matrix of the nonergodicity parameters ${\bf f}({\bf q})$ for the
coherent intermediate scattering function is given by the solution of
Eq.~(\ref{eq1}) at long times, i.e. $f_{ij}({\bf q})=lim_{t\to \infty}
F_{ij}({\bf q},t)$. It can be shown that ${\bf f}({\bf q})$ can be
computed via the following iterative procedure~\cite{fuchs_phd}:
\begin{equation}
{\bf f}^{(l+1)}(q) = \frac{
{\bf S}(q) \cdot {\bf N}[{\bf f}^{(l)},{\bf f}^{(l)}](q) \cdot {\bf S}
(q) + q^{-2}|{\bf S}(q)| |{\bf N}[{\bf f}^{(l)},{\bf f}^{(l)}](q)| 
{\bf S}(q)
                           }{
q^2+Tr({\bf S}(q) \cdot {\bf N}[{\bf f}^{(l)},{\bf f}^{(l)}](q)) + 
q^{-2}| {\bf S}(q)| | {\bf N}[{\bf f}^{(l)},{\bf f}^{(l)}](q)|}
\quad,
\label{eq9}
\end{equation}
where the matrix ${\bf N}(q)$ is given by
\begin{equation}
N_{ij}(q)=\frac{m_i}{x_i k_B T} M_{ij}(q) \quad.
\label{eq10}
\end{equation} 

For temperatures above the critical temperature $T_c$ this iteration
converges to the trivial solution ${\bf f}(q)=0$ whereas for $T<T_c$
it converges to a nontrivial solution ${\bf f}(q)>0$.

The incoherent nonergodicity parameter $f_i^{(s)}$ can be computed from the
following iterative procedure:
\begin{equation}
q^2 \frac{f_i^{(s,l+1)}(q)}{1-f_i^{(s,l+1)}(q)} = M_i^{(s)}[{\bf f},
f_i^{(s,l)}](q)
\quad .
\label{eq11}
\end{equation}

In order to determine the critical point it is useful to consider the
so-called stability matrix ${\bf C}$ which is defined by its action
on a vector $\delta {\bf f}(q)=(\delta {\bf f}_{11}(q),
\delta {\bf f}_{12}(q), \delta {\bf f}_{22}(q))$:
\begin{equation}
({\bf C}\cdot \delta {\bf f})(q)=\frac{1}{q^2} ({\bf S}(q)-{\bf f}(q))
\cdot \left[ {\bf M}[{\bf f},\delta {\bf f}](q) + {\bf M}[\delta {\bf
f},{\bf f}](q) \right] \cdot
({\bf S}(q)-{\bf f}(q)) \quad .
\label{eq12}
\end{equation}

We define $E_0$ to be the largest eigenvalue of this matrix, ${\bf
e}(k) = (e_{11}(k),e_{12}(k),e_{22}(k))$ as the corresponding right and
$\hat{\bf e}(k)=(\hat{e}_{11}(k),\hat{e}_{12}(k),\hat{e}_{22}(k))$ 
as the corresponding
left eigenvector of this matrix. The normalization of these
eigenvectors is given by~\cite{fuchs_priv}:
\begin{eqnarray}
\int_0^{\infty} dk \sum_{n=11,12,22} \hat{e}_{n}(k) e_n(k) & = & 1
\label{eq13a} \\
\int_0^{\infty} dk \sum_{n=11,12,22} \hat{e}_n(k)\left[{\bf e}(k)\cdot \left[
{\bf S}(k)-{\bf f}(k) \right]^{-1}\cdot {\bf e}(k)\right]_n & = & 1
\quad .
\label{eq13b}
\end{eqnarray}

The critical amplitudes ${\bf h}(q)=(h_{11}(q),h_{12}(q),h_{22}(q))$
describe the dynamics of the system in the $\beta$-relaxation regime,
i.e.
\begin{equation}
F_{ij}(q,t)=f_{c,ij}(q)+h_{ij}(q)g(t)\quad ,
\label{eq14}
\end{equation}
where $g(t)$ is a function which is independent of $q$, and whose form
depends on temperature and the so-called exponent parameter $\lambda$,
and $f_{c,ij}$ are the nonergodicity parameters at the critical
temperature. This critical amplitude is given by the value of the right
eigenvector at the critical temperature $T_c$, i.e.:
\begin{equation}
h_{ij}(q)=e_{c,ij}(q)\quad .
\label{eq15}
\end{equation}

The value of the mentioned exponent parameter $\lambda$ is given by:
\begin{equation}
\lambda = \int_0^{\infty} dq \sum_{n=11,12,22} \hat{e}_{c,n}(q) 
\left(\frac{1}{q^2}({\bf S}(q)-{\bf f}_c(q))\cdot {\bf M}[{\bf e}_c,
{\bf e}_c]\cdot ({\bf S}(q)-{\bf f}_c(q))\right)_{n}
\label{eq16}
\end{equation}

The procedure to compute the nonergodicity parameters is now the
following: Given the partial structure factors for a temperature $T$ which
corresponds to the glass state one computes from Eqs.~(\ref{eq5}) and
(\ref{eq6}) the memory kernel and iterates Eq.~(\ref{eq9}) until ${\bf
f}^{(l)}(q)$ has converged. Then the stability matrix ${\bf C}(q)$ and
its largest eigenvalue $E_0$ are computed. It can be shown that in the
vicinity of the critical temperature $T_c$ the relation
\begin{equation}
(1-E_0)^2= A (T_c-T) +O((T_c-T)^2)
\label{eq17}
\end{equation}
holds, which can thus be used for a precise determination of $T_c$.
Having determined $T_c$ we can compute the right and left eigenvalue of
${\bf C}$ at $T_c$ and thus obtain the critical amplitudes $h_{ij}(q)$
and the exponent parameter $\lambda$ (Eqs.~(\ref{eq15}) and
(\ref{eq16})). Using the nonergodicity parameters of the coherent
intermediate scattering function we can use Eqs.~(\ref{eq7}) and
(\ref{eq11}) to finally compute the nonergodicity parameter for the
incoherent intermediate scattering function.

\section{Model and Computational Details}
\label{sec3}

In this section we introduce the system we investigate and give some of 
the details of our numerical calculations. More details on these
calculations can be found in Ref.~\cite{nauroth95}.

The model we are studying is a binary mixture of classical
Lennard-Jones particles all of them having mass $m$. The interaction
between two particles of type $i$ and $j$ ($i,j \in \{A,B\})$ is given
by $V_{ij}(r)=4\epsilon_{ij}\left[ (\sigma_{ij}/r)^{12} -
(\sigma_{ij}/r)^6\right]$. The parameters $\epsilon_{\alpha\beta}$ and
$\sigma_{\alpha\beta}$ are given by: $\epsilon_{AA}=1.0$,
$\epsilon_{AB}=1.5$, $\epsilon_{BB}=0.5$, $\sigma_{AA}=1.0$,
$\sigma_{AB}=0.8$ and $\sigma_{BB}=0.88$. The composition of the
mixture is such that $x_A=0.8$ and $x_B=0.2$. In the following we will
measure length scales in units of $\sigma_{AA}$ and energy in units of
$\epsilon_{AA}$ and set $k_B=1$.

In a recent simulation the dynamics of this system was investigated by
means of a molecular dynamics computer
simulation~\cite{kob94,kob95a,kob95b,kob95c,kob95d}. This simulation
used 800 particles of type $A$ and 200 particles of type $B$. The size of
the cubic box was held fixed at $L=9.4$. In order to lower
the computational burden the Lennard-Jones potential was truncated and
shifted at a distance of $2.5\sigma_{\alpha\beta}$. More details on
that simulation can be found in the original papers. In that work also
the partial structure factors $S_{ij}(q)$ were calculated. This was
done by computing the space Fourier transform of the radial
distribution function $g_{ij}(r)$. Because of the finite size of the
system this Fourier transform gave rise to unphysical oscillations in
the structure factors at small values of $q$. Since these structure
factors are the (only) input in the mode-coupling equations, such
unphysical oscillations would possibly modify the outcome of the MCT
calculations. Therefore we repeated the simulations and computed the
partial structure factor directly from the positions of the particles
by means of Eq.~(\ref{eq0}) and thus avoided the above mentioned
Fourier transform. In order to filter out high frequency noise in $q$
the so determined structure factors were smoothed with a spline under
tension.  Because of the finite size of the box, wave vectors with
modulus less than $2\pi/L$ are not accessible.  Therefore we
extrapolated the determined partial structure factors to $q=0$.

These new simulations were done only for a few selected values of the
temperature, all of them in the vicinity of the critical temperature,
i.e. at $T=1.0$, $T=0.8$ and $T=0.6$. For a precise determination of
the critical temperature we also needed the structure factors at
intermediate values of the temperature. Therefore we used the structure
factors at the three mentioned temperatures and a quadratic
interpolation scheme to compute the structure factors for intermediate
temperatures. Since in this temperature region the structure factors
show only a weak variation with temperature such an interpolation
scheme should be fairly reliable.  More details on the so obtained
structure factors are given in the next section.

The integral equations presented in the previous section were solved
iteratively in the way described in that section. The occurring
integrals were computed using a high order Simpson scheme. (Note that
it is necessary to use an integration scheme that accesses only points
that are spaced in an equidistant way since the integrals involve
convolutions. Therefore more efficient integration schemes like
Gaussian quadrature cannot be used.) The upper limit of the integrals
was set to $q_{co}=40$, which is sufficiently large to allow the direct
correlation function to be negligible small for $q>q_{co}$.  In order
to perform the integration we used 300 grid points in the interval
$[0,q_{co}]$. A few calculations with a larger number of points showed
that this number is sufficiently large to neglect the dependence of the
final results on the used discretisation scheme.

Close to the glass transition the convergence of the iteration scheme
given by Eq.~(\ref{eq9}) was quite slow, since the maximum eigenvalue
was very close to unity ($1-E_0\approx 3\cdot 10^{-3}$, which
corresponds to $T-T_c \approx 1\cdot 10^{-5}$), and only after
1000-2000 iterations reliable results could be obtained. Thus such an
iteration took about 24 hrs on a medium level workstation. Note that in
order to get results that are accurate to within one percent it is
indeed necessary to determine $T_c$ that precisely, since, e.g.,
quantities like the nonergodicity parameters show a square root
dependence on $(T-T_c)$.

\section{Results}
\label{sec4}

In this section we present the obtained results. In the first part we
investigate whether MCT, as presented in Sec~\ref{sec2}, is able to
predict correctly various quantities that are relevant in the dynamics
of the supercooled liquid. In the second part we test whether two
possible modifications of MCT lead to an even better agreement
between theory and the simulation.

The partial structure factors $S_{ij}(q)$, crucial input for the
theory, were computed as described in Sec.~\ref{sec3}. The resulting
structure factors are shown in Fig.~\ref{fig1} for $T=1.0$, $T=0.8$ and
$T=0.6$. (Note that, although in our computation we used $S_{ij}(q)$
for all values of $q$ up to $q_{co}=40$ we show only the range $0\leq
q \leq 20$, since outside this interval the structure factors are almost
constant.) From this figures we see that in that range of temperature
the dependence of the structure factors on temperature is very smooth,
thus justifying the interpolation procedure described in
Sec.~\ref{sec3} to obtain the structure factors at intermediate values
of $T$.

Using the procedure described in Sec.~\ref{sec2}, we determined the
critical temperature $T_c$ to be around 0.922. This value has to be
compared with the result that was obtained from the molecular dynamics
computer simulation, which was
$T=0.435\pm0.003$~\cite{kob94,kob95a,kob95b,kob95c,kob95d}. Thus we
find that, for the system considered here, MCT overestimates the
critical temperature by about a factor of two.  Since the idealized
theory neglects certain types of relaxation processes, which are
usually called hopping processes, it can be expected that the critical
temperature predicted by the theory is too high. Nevertheless, {\it at
first sight}, the factor of two seems to be surprisingly large when
compared with the results of similar comparisons between the prediction
of MCT for the value of the critical coupling parameter and the results of
experiments or computer simulations. E.g. it was found in light
scattering experiments on colloidal particles, a system which is
considered to be described well by a hard sphere model, that these
systems undergo a glass transition at a packing density $\phi_c$ which
is between 0.56 and 0.58~\cite{colloids}. This value compares well with
the critical packing density of $0.52\pm0.01$ of MCT for a system of
hard spheres~\cite{bgs}. Thus in this case the discrepancy between the
experiments and the theory is about 10\%. In the case of a binary
system of soft spheres, i.e. a pair interaction which is proportional
to $r^{-12}$, it was found by means of computer simulations that the
glass transition occurs at a value of the effective coupling constant
$\Gamma$ of 1.46~\cite{bernu87,roux89}. It can be shown that $\Gamma$
is the only relevant parameter for the thermodynamic state of such a
system and that
\begin{equation}
\Gamma \propto nT^{-1/4},
\label{eq18}
\end{equation}
where $n$ is the particle density.  Using the Roger-Young integral
equations to compute the structure factors for this system, Barrat and
Latz computed the critical value of the coupling constant within the
framework of MCT and found it to be 1.32~\cite{barrat90}. Using
expression~(\ref{eq18}) 
we thus find that the discrepancy is about 10\% in the
density, comparable to the above mentioned discrepancy for the hard
sphere system, but about 50\% in the critical temperature.

Expression~(\ref{eq18}) is valid only for a soft sphere system.
However, since at low temperatures it is mainly the repulsive core of
the particles that is important for the structure of the liquid, it can
be expected that for a Lennard-Jones system, which has the same type of
hard core as the soft sphere system, expression~(\ref{eq18}) is a
reasonable approximation. This expectation is corroborated by the
calculations of Bengtzelius in which the critical temperature $T_c$ was
determined at different densities for the case of a one component
Lennard-Jones system (the structure factor was computed from the
so-called optimized random-phase approximation)~\cite{bengtzelius86}.
In that work it was shown that a change of 10\% in density gives rise
to a change of a factor of two in $T_c$, in qualitative agreement with
the results for the soft sphere system. Thus we conclude that a
discrepancy of a factor of two in the critical temperature correspond
to a discrepancy between theory and simulation of only 20\% in the
coupling constant, which is comparable to the discrepancy found in the
above mentioned hard sphere system and the soft sphere system.

The next quantity for which we compare the prediction of the theory
with the result of the computer simulation is the exponent parameter
$\lambda$. In the simulation this parameter was determined by fitting
the functional form provided by MCT for the
$\beta$-correlator~\cite{bibles} to the corresponding master curves
found in the $\beta$-relaxation regime~\cite{kob95b}.  Depending on the
type of correlator investigated, the value of $\lambda$ was found to
vary between 0.75 and 0.83 with the most likely value of
$\lambda=0.78\pm 0.02$.  Our MCT calculations showed that for this
system the theory predicts a value of 0.708, which compares quite well
with the one found previously in the simulation.  The discrepancy in
$\lambda$ between theory and computer simulation is in our case smaller
than the one found for the soft sphere system, for which the theory
predicted $\lambda$=0.73~\cite{barrat90} and for which $\lambda\approx
0.61$ was found in the simulation.  This latter value has, however,
probably a relatively large error bar, since it was determined from the
critical exponent of the diffusion constant close to $\Gamma_c$.  This
critical exponent was determined to be around 2.0~\cite{bernu87,roux89}
and, if it is assumed that the connection predicted by MCT between this
critical exponent and the exponent parameter $\lambda$
holds~\cite{bibles}, one obtains the quoted value of $\lambda$.
However, there are two reasons why the so determined value of $\lambda$
might be slightly wrong: First the value of the critical exponent is
not known very precisely~\cite{bernu87,roux89} and secondly it might
well be that the above mentioned connection between the critical
exponent for the diffusion constant and the exponent parameter holds
only very close to the critical temperature (or coupling parameter).
E.g., in the simulation of the binary Lennard-Jones system it has been
found that the critical exponent for the diffusion constant and the one
for the $\alpha$-relaxation times differ by about 20\%~\cite{kob95b},
although MCT predicts these exponents to be the same. Thus this shows
that either the connection proposed by MCT is not always valid or that,
in order to see the correct critical behavior, one has to be much
closer to the critical temperature than it is currently possible with
computer simulations (because very close to the critical point the time
scales of the $\alpha$-relaxation exceeds the time scale accessible to
such simulations). Furthermore it can be that in the vicinity of $T_c$
the so-called hopping processes become important for the soft sphere
system and thus give rise to an {\it effective} exponent which is
different from the one predicted by the theory.  Thus because this
connection between the critical exponent of the diffusion constant and
the exponent parameter $\lambda$ is not beyond any doubt, the value of
$\lambda$ in the soft sphere model is not known very precisely. Finally
we mention that in the comparisons between the prediction of MCT and
the results of the experiments on colloidal systems the latter were
always assumed to be hard spheres and thus the exponent parameter was
not a fit parameter~\cite{goetze91}.

We now turn our attention to the wave-vector dependence of the
nonergodicity parameter. This quantity was determined in the
simulation~\cite{kob95b} and the results are shown for the coherent
intermediate scattering functions for the $AA$, the $AB$ and the $BB$
correlation (bold dashed lines), as well as for the incoherent
intermediate scattering function for the $A$ and the $B$ particles in
Fig.~\ref{fig2} (thin dashed lines). For small values and very large
values of $q$ it was not possible to determine $f_c$ from the
simulation because of problems with finite size effects and statistics.
Also included in the figure are the predictions of MCT (solid lines).
First we consider Fig.~\ref{fig2}a which shows $f_c(q)$ for the $AA$
correlation and the $A$ particles. The first observation is that the
theoretical curves match the ones of the simulations qualitatively well
for all values of $q$ in that the location of the various extrema in
the $f_c(q)$ for the coherent scattering function are reproduced
correctly.  In addition also the fact that for large values of $q$ the
nonergodicity parameter for the coherent scattering function oscillates
around the one for the incoherent one is reproduced correctly by the
theory.

For wave-vectors in the vicinity of the first peak of the structure
factor, also the {\it quantitative} accordance between theory and
simulation is very good. This agreement is less good for wave-vectors
larger than the second peak in the structure factor. This can be due to
two reasons: The first one is that the nonergodicity parameters as
determined from the simulation are affected by systematic errors of
unknown magnitude~\cite{kob95b}. From the way these quantities were
measured it can be expected that these errors increase with increasing
wave-vector which might be the reason for the increasing discrepancy
between simulation and theory. The second possible reason is that for
large values of $q$ MCT is no longer reliable.  This can be understood
as follows: The large wave-vectors correspond to distances which are
relatively small compared to the diameter of the particles.  Now one
should remember that in the derivation of the MCT equations a
factorization ansatz was made. This ansatz is reasonable for distances
on the order of the diameter of a particle but is likely to be bad for
much smaller distances. Thus it is expected that the vertices
$V_{i\alpha\beta}$ of Eq.~(\ref{eq4}) are not quite correct for large
values of $q$ or $k$. Therefore it is not surprising that the
accordance between the results of the simulations and the predictions
of MCT is not as good for large values of $q$ as it is for wave-vectors
in the vicinity of the peak of the structure factor.

Furthermore we comment on two smaller features in the curves. First we
see that the curve for the nonergodicity parameter for the coherent
scattering functions, as computed from the simulation, shows to the left
and to the right of the large peaks (at $q\approx 7$ and $q\approx 12$)
a small peak. These small peaks are a finite size effect which is due
to the method we computed the intermediate scattering
function~\cite{kob95b}. Thus the fact that these small peaks are not
present at all in the curve as computed from MCT should not be viewed
as a failure of the theory to reproduce this feature. Secondly we see
that the MCT curve for the coherent intermediate scattering function
shows some small peaks for wave-vectors smaller than 2.  This behavior
is likely to be due to numerical instabilities in the computation of
this curve and therefore has no physical relevance.

The wave-vector dependence of the nonergodicity parameter for the $AB$
correlation is shown in Fig.~\ref{fig2}b.  In order to make clear where
the measured points actually are, we show them as open squares and the
connecting dashed line should be considered just as a guide to the eye.
We recognize that this $q$-dependence is very different from the one
found for the $AA$ correlation. We see that for values of $q$ near
$q=7$ and near $q=10$ there is a gap in the data of the simulation. The
reason for this is that in the vicinity of these wave-vectors the
partial structure factor of the $AB$ correlation changes sign which in
turn leads to a singularity and hence to numerical difficulties in
determining the corresponding intermediate scattering function.  Also
included in the figure is the prediction of MCT for this nonergodicity
parameter. We see that for intermediate values of $q$ the theoretical
curve describes the data very well and we see that MCT correctly
predicts the presence of the just mentioned singular behavior of the
nonergodicity parameter.  For larger values of $q$ the agreement is only
qualitatively correct and the probable reason for this has been given
above.

In Fig.~\ref{fig2}c we show the wave-vector dependence of the
nonergodicity parameter for the coherent and incoherent intermediate
scattering function for the $B$ particles. From this figure we see that
for the coherent part this dependence is very different from the one
for the $A$ particles in that it resembles much more the $q$-dependence
of the incoherent part. This can be qualitatively understood by
remembering that the number of $B$ particles is smaller by a factor of 5
than the number of $A$ particles. Thus since the $B$ particles are
relatively dilute the coherent intermediate scattering function behaves
very similar to the single particle correlation function, i.e. the
incoherent intermediate scattering function.

As we can see from the figure, MCT is able to reproduce the wave-vector
dependence also of these two nonergodicity parameters very well.
Although the theoretical curves lie below the ones from the computer
simulation for all values of $q$, the agreement is nevertheless on the
order of about 5\% for intermediate values of $q$.

From figure~\ref{fig2} we recognize that in all cases the
nonergodicity parameters as determined from the simulations is a bit
larger than the ones predicted by MCT. This is in qualitative agreement
with our observation that the critical temperature as found in the
simulation is quite a bit lower than the one predicted by the theory,
which, in a hand wavy fashion, can be understood as follows:  The
nonergodicity parameter is some sort of measure for how much a
particle can move in the cage formed by its surrounding particles.
Since it can be expected that this movement is smaller the lower the
temperature is, it follows that the nonergodicity parameter increases
with decreasing temperature. Thus we see that if MCT would have
predicted a critical temperature which is lower than the one it
predicts now, we would expect that the theoretical nonergodicity
parameters would be larger than the ones the theory predicts now.  Thus
we have evidence that the too high critical temperature and the
nonergodicity parameters that are too small are related phenomena. We
will come back to this point below.

The fact that the predicted nonergodicity parameters are smaller than
the ones determined from the simulation animated us to
compare the former with the amplitude of the $\alpha$-relaxation.
It should be remembered that in the simulation the nonergodicity
parameter was determined from the height of the plateau in the time
correlation function. In these simulations it was shown that the height
of this plateau is {\it not} equivalent to the amplitude of the
Kohlrausch-Williams-Watts (KWW) function which describes the relaxation on
time scales beyond the $\beta$-relaxation time scale, i.e.  $\phi(t)=
A\exp(-(t/\tau)^{\beta})$. Since we have found that this KWW-amplitude
$A$ is always a bit smaller than the nonergodicity
parameter~\cite{kob95b,kob95d}, it is interesting to compare the
$q$-dependence of this measured amplitude with the $q$-dependence of
the nonergodicity parameter as predicted by MCT. This is done in
Fig.~\ref{fig3}. We see that for all correlation functions the
agreement between these two quantities is very good. To our surprise we
find that this agreement is always better than the one between the
nonergodicity parameter of the simulation and the one of MCT, which was
shown in Fig.~\ref{fig2}. (An exception is the $AB$ correlation in the
range $5\leq q \leq 7$, where the experimental point for $A$ are now
clearly below the MCT curve.) At the moment it is not clear to us
whether this surprising accordance between the KWW-amplitude and the
nonergodicity parameter of MCT is just a coincidence or whether there
is some underlying reason for it. One possibility might be that the
corrections to the asymptotic scaling laws of MCT are larger for the
$\beta$-relaxation regime than for the $\alpha$-relaxation regime. Thus
it would be interesting to compute the full time dependence of the
correlation functions within the framework of MCT and to compare the so
obtained results with the results from the simulations. In addition it
would be helpful to make similar comparisons with other systems in
order to see whether the just described phenomenon holds for other
systems as well.

The last quantity we investigate is the wave-vector dependence of the
critical amplitudes $h(q)$. These amplitudes are used to describe the
time dependence of a correlation function in the $\beta$-relaxation
regime, see Eq.~(\ref{eq14}).  In the computer simulation it was found
that in the $\beta$-relaxation regime the various intermediate
scattering functions are indeed of the form of
Eq.~(\ref{eq14})~\cite{kob95a} in that it was demonstrated that the left
hand side of
\begin{equation}
\frac{\Phi(r,t)-\Phi(r,t')}{\Phi(r',t)-\Phi(r',t')}=\frac{H(r)}{H(r')}
\quad,
\label{eq20}
\end{equation}
which is obtained from the space Fourier transform of Eq.~(\ref{eq14}),
holds if the times $t$ and $t'$ are on the time scale of the
$\beta$-relaxation. Here $r'$ can be chosen arbitrarily. From that
calculation it was possible to estimate an upper and lower bound for
$H(r)$, and these bounds are shown in Fig.~\ref{fig4} for the three
coherent intermediate scattering functions (thin solid lines). The
value of $r'$ was 1.095, 0.9 and 1.73 for the $AA$, the $AB$ and the $BB$
correlation. Also included in the figures are the prediction of MCT for
these quantities (bold solid lines). From these figures we recognize that
the agreement between the results of the computer simulation and the
prediction of the theory is, in the case of the $AA$ and the $AB$
correlation, qualitatively as well as quantitatively very good in that
also small details in the curves, such as the small dip in the peak at
around $1.8$, are reproduced correctly.  The agreement between
simulation and theory is not that good for the case of the $BB$
correlation in that the amplitude of the various peaks is not predicted
correctly. However, the position of these peaks is in accordance with
the theory and thus MCT is correct for this correlation function at
least qualitatively. We also notice that in all three cases the
agreement between theory and simulation is not very good at small
distances. This is not surprising, since we have already explained above
why MCT is not very accurate for large wave-vectors, i.e. small
distances.

In order to gain some insight into the nature of the various peaks in
$H(r)$ we have included in the figures also the corresponding radial
distribution functions $g_{ij}(r)$ at $T=0.466$, the lowest temperature
considered in the computer simulation~\cite{kob95a} (dashed curves). We
see that, for values of $r$ larger than the first nearest neighbor
peak, the different maxima and minima in $H(r)$ occur at the same
location at which the corresponding $g_{ij}(r)$ shows its extrema. This
means that in $q$-space $h_{ij}(q)$ shows extrema at the same values
of $q$ as the corresponding partial structure factor. Since the latter
shows a similar $q$-dependence as the corresponding coherent
nonergodicity parameter, see Fig.~\ref{fig2}, the nonergodicity
parameter and $h_{ij}(q)$ will show extrema at the same values of $q$.
A similar observation was made in the case of the soft sphere
system~\cite{barrat90} and thus it can be conjectured that this is a
general rule.

In the remaining of this section we present the results we obtained by
considering two small modification of MCT. These modifications were
done in an attempt to improve the agreement between the results of the
computer simulation and the prediction of the theory.

The basic idea of the first modification is as follows. In the first
part of this section we have shown that MCT is able to give a
surprisingly good description of the $q$-dependence of the
nonergodicity parameters and the critical amplitudes. The most severe
disagreement seems to be that the theory overestimates the critical
temperature by quite a bit. Furthermore we have seen that the theory
underestimates the nonergodicity parameter and that this disagreement
is most pronounced at large values of $q$. In the discussion of this
effect we argued that one of the reason for its occurrence might be
that a factorization ansatz, which is used in the derivation of the
MCT equations, breaks down for small distances and that therefore the
MCT equations are not accurate for these values of $q$.  Therefore one
could argue that it is better to leave out from the calculation of the
memory kernel in Eq.~(\ref{eq3}) that part of the wave-vector
integration altogether, i.e. to restrict the integration to
wave-vectors with modulus less than a certain limit $q_{co}$. This
approximation is equivalent to the assumption that the structure factor
is constant for $q\geq q_{co}$. Thus the value of $q_{co}$ can be
used as a fit parameter in order to match the critical temperature as
predicted by MCT with the one determined from the simulation. The hope
is that this fix of the critical temperature will lead to theoretical
nonergodicity parameters that are in better agreement with the ones of
the simulation (of course at the cost that for wave-vectors larger than
$q_{co}$ the theory does not give any nonergodicity parameter at
all).

Thus we proceeded as follows. Using the partial structure factors that we
determined at $T=0.466$, $T=0.475$ and $T=0.5$ we made an extrapolation
to determine the partial structure factors at $T=0.435$, the value of the
critical temperature as determined from the computer simulation. Since
at these low temperatures the temperature dependence of the structure
factors is only weak and very regular, such an extrapolation is not
problematic. Equipped with the structure factors at the correct
critical temperature we determined $q_{co}$ such that the critical
temperature as determined from MCT is exactly at $T=0.435$, i.e. the
critical temperature from the simulation. The value of $q_{co}$ we
obtained was around 11.7, i.e.  a bit to the right of the first minimum
in the partial structure factor for the $AA$ correlation. This value
shows on the one hand that it is mainly the first peak in the structure
factor that is relevant to give a {\it qualitative} correct description
of the transition and on the other hand that for a {\it quantitative}
correct calculation of the transition temperature it is necessary to
take into account the structure also for larger values of $q$. With
this value of $q_{co}$ we computed the $q$-dependence of the
nonergodicity parameters and the result for the $AA$ correlation are
shown in Fig.~\ref{fig5} (solid line). Also included is the curve from
the simulation (bold dashed line) and, as a reference, the curve when
$q_{co}$ is 40, i.e. the value of $q_{co}$ used to compute the
results of the first part of this section (thin dashed line). From this
figure we recognize that the curve for the new value of $q_{co}$ is
now significantly below the curve from the simulation and that the
discrepancy between the (modified) theory and simulation is now quite a
bit larger than it was with the original theory. Thus this shows two
things: First that the contribution to the memory kernel that come from
values of $q$ larger than $q_{co}=11.7$ are important in order to get
quantitatively correct results, despite the above discussed fact that
the integrand is not quite appropriate for such large values of $q$,
and secondly that it is not that easy to improve the theory
qualitatively by introducing a fit parameter in order to fix certain
shortcomings of the theory (as e.g. the not so satisfactory prediction
of the critical temperature). 

The second modification of the theory we did was to ignore the fact
that we have to compare the results for the $q$-dependence of the
nonergodicity parameter as obtained from the simulation with the
prediction of MCT for the nonergodicity parameter {\it at the critical
temperature $T_c$}. Since we have seen that MCT underestimates the
nonergodicity parameters and we know that for temperatures $T<T_c$ the
theory predicts that the nonergodicity parameters increase, we tried to
correct this discrepancy by computing $f_A^{(s)}(q)$ at a temperature
$T_{eff}$ below $T_c$ and to determine $T_{eff}$ by requiring that this
nonergodicity parameter, which we will call $f_{eff}^{(s)}$, fits the
corresponding quantity from the simulation well. The reason for
choosing this type of nonergodicity parameter, instead of e.g. the one
for the $AA$ correlation, is that in the simulation
it can be determined with the best accuracy. This fit gave a value for
$T_{eff}$ around 0.91, thus quite close to the original critical
temperature $T_c=0.922$. The resulting $q$-dependence of
$f_{eff}^{(s)}$ is shown in Fig.~\ref{fig6}a together with $f_c^{(s)}$
from the simulation. We see that for small and intermediate values of
$q$ the agreement between $f_{eff}^{(s)}$ and $f_c^{(s)}$ is very good.
Only for large values of $q$ significant, but not large, discrepancies
occur. We also computed the $q$ dependence of the other nonergodicity
parameters for the {\it same temperature $T_{eff}$} and the results are
shown in Fig.~\ref{fig6}. From Fig.~\ref{fig6}a we see that for the
coherent nonergodicity parameter for the $AA$ correlation the agreement
between $f_c(q)$ and $f_{eff}(q)$ has improved significantly compared
to the agreement when the original MCT function is used (see
Fig.~\ref{fig2}a) and that for $q$ values in the vicinity of the first
peak and the first maximum the agreement is perfect to within the noise
of the simulation data. Also for the $AB$ correlation function
(Fig.~\ref{fig6}b) the agreement between simulation and theory has
improved considerably compared to the original MCT and the same
conclusion holds for the coherent and incoherent nonergodicity
parameters for the $B$ particles (Fig.~\ref{fig6}c). Thus we conclude
that introducing one fit parameter, namely the temperature $T_{eff}$
at which the nonergodicity parameters are evaluated within the
framework of MCT leads to a significant improvement of the agreement
between theory and simulation.

\section{Summary and Discussion}
\label{sec5}

We have presented the results of a numerical calculation in which the
mode-coupling equations were solved for a binary Lennard-Jones
mixture.  The goal of these calculations was to test whether the
agreement between the predictions of MCT for the dynamics, which was
investigated by means of a computer
simulation~\cite{kob94,kob95a,kob95b,kob95c,kob95d}, holds only for the
universal predictions of the theory or also for the nonuniversal ones.
Using the partial structure factors, as determined from a computer
simulation, as input, we computed within the framework of MCT the
critical temperature, the exponent parameter, the $q$-dependence of the
various nonergodicity parameters and the various critical amplitudes.
Although the critical temperature as predicted by MCT is a factor of
two larger than the one determined from the computer simulation, we
argue that this discrepancy is significantly smaller when expressed
through the effective coupling constants, and is then comparable with
the discrepancies found for this quantity for systems like hard
spheres~\cite{colloids} or soft spheres~\cite{barrat90}. The exponent
parameter as predicted by the theory is in fair agreement with the one
determined from the simulation. MCT makes a very good quantitative
prediction for the wave-vector dependence of the nonergodicity
parameter for values of $q$ in the vicinity of the first maximum and
the first minimum. For large values of $q$ the agreement is still fair
and we can rationalize the increasing discrepancy between theory and
simulation by arguing that some of the approximations used to derive
the mode-coupling equations no longer hold in this limit.  In addition
we also showed that the theory is also able to make a quantitatively
correct prediction of the various critical amplitudes $H(r)$.

We also compared the $q$-dependence of the Kohlrausch-Williams-Watts
amplitude, as determined from the simulation, with the $q$-dependence of
the nonergodicity parameter as predicted by MCT and found that the two
quantities match surprisingly well. So far it is not clear why this is
the case and how general this observation is. Therefore it would be
very valuable to make similar comparisons for different types of
systems. 

In an attempt to improve the agreement between the measured and
theoretical nonergodicity parameters we introduced an upper cut-off
$q_{co}$ in the integral of the memory kernel and used $q_{co}$ to
match the critical temperature $T_c$ between MCT and simulation.  We
found that the introduction of this fit parameter leads to a worsening
of the agreement between the measured and theoretical nonergodicity
parameter which shows that the contributions to the memory kernel from
large values of $q$ are important for a quantitative correct
description of the nonergodicity parameter. In a second
``modification'' of the theory we used temperature as a fit parameter
and determined a temperature $T_{eff}<T_c$ such that the resulting
incoherent nonergodicity parameter for the $A$ particles fits the
simulation data well. We found that at this temperature also all the
other nonergodicity parameters fit the data from the simulation well,
in some cases even very well.  Thus it seems that there exists a
temperature $T_{eff}$ for which MCT is able to predict very well the
$q$-dependence of the various nonergodicity parameters as measured in
the simulation {\it at $T_c$}.  This shows that the intrinsic structure
of the mode-coupling equations are clearly able to correctly describe
such quantities and that it is perhaps only through the omission of
certain contributions to the memory-kernel that there is no perfect
quantitative agreement between the prediction of the theory and the
results of the simulation.

To summarize we can say that our calculations have shown that
MCT is able to give a correct {\it quantitative} description of the
dynamics of a simple liquid if one restricts oneself to quantities
like the critical temperature, the exponent parameter, the
nonergodicity parameter or the critical amplitudes. Whether MCT is also
able to give a correct description of the {\it full} time dependence of
the various correlation functions, as it is the case for the
$\beta$-relaxation regime in colloidal systems~\cite{goetze91}, remains
to be tested and is subject of ongoing work.

Acknowledgements: We thank M. Fuchs for extensive help and enlightening
discussions during this work, W. G\"otze for valuable discussions and
useful comments on the manuscript, K. Binder for helpful comments on
the manuscript and J.L. Barrat for providing us with
some programs which allowed to check our programs. This work was
supported by SFB 262/D1 of the Deutsche Forschungsgemeinschaft.

\newpage

\begin{figure}
\caption{Partial structure factors for those temperatures used for
the interpolation to obtain the partial structure factors at the
critical temperature. $T=1.0$ (dotted line) $T=0.8$ (dashed line) and
$T=0.6$ (solid line). a) $AA$-correlation b) $AB$-correlation c)
$BB$-correlation.
\protect\label{fig1}}
\vspace*{5mm}
\par

\caption{Nonergodicity parameter $f_c(q)$ for the coherent (bold lines)
and incoherent (thin lines) intermediate scattering functions as
determined from the simulation (dashed lines) and as predicted by MCT
(solid lines). a) $A$ particles and $AA$ correlation, b) $AB$ correlation c)
$B$ particles and $BB$ correlation.
\protect\label{fig2}}
\vspace*{5mm}
\par

\caption{Kohlrausch-Williams-Watts amplitude $A$ for the coherent (bold
lines) and incoherent (thin lines) intermediate scattering functions as
determined from the simulation (dashed lines) and nonergodicity
parameter as predicted by MCT (solid lines). a) $A$ particles and $AA$
correlation, b) $AB$ correlation c) $B$ particles and $BB$ correlation.
\protect\label{fig3}}
\vspace*{5mm}
\par

\caption{Critical amplitude $H(r)$ for the coherent intermediate
scattering function as predicted by MCT (bold solid line) and the upper
and lower bound for this function (thin lines) as determined from the
simulation. The dashed lines are $x$ times the corresponding radial
distribution function at $T=0.466$. a) $AA$ correlation $x=0.25$, b) 
$AB$ correlation $x=0.2$, c) $BB$ correlation $x=0.8$.
\protect\label{fig4}}
\vspace*{5mm}
\par

\caption{Nonergodicity parameter for the coherent intermediate
scattering function for the $AA$ correlation as determined from the
simulation (bold dashed line) and the prediction of MCT for
$q_{co}=11.7$ (solid line) and $q_{co}=40$ (dashed line).
\protect\label{fig5}}
\vspace*{5mm}
\par

\caption{Nonergodicity parameter $f_c(q)$ for the coherent (bold lines)
and incoherent (thin lines) intermediate scattering functions as
determined from the simulation (dashed lines). The solid lines are
results for the nonergodicity parameter as predicted by MCT for the
temperature $T_{eff}=0.91$. a) $A$ particles and $AA$ correlation, b) $AB$
correlation c) $B$ particles and $BB$ correlation.  
\protect\label{fig6}}
\vspace*{5mm}
\par

\end{figure}

\begin{references}
%
\bibitem[*]{wkob}
Author to whom correspondence should be addressed to.
Electronic mail: kob@moses.physik.uni-mainz.de;\\
http://www.cond-mat.physik.uni-mainz.de/\~{ }kob/home\_kob.html
%
\bibitem{bibles}
W. G\"otze, p. 287 in {\it Liquids, Freezing and the Glass
Transition} Eds.: J. P.  Hansen, D. Levesque and J. Zinn-Justin, 
Les Houches.  Session LI, 1989, (North-Holland, Amsterdam, 1991);
%
W. G\"otze and L.  Sj\"ogren, Rep. Prog. Phys. {\bf 55}, 241 (1992).
%
\bibitem{schilling92}
R. Schilling, p. 193 in {\it Disorder Effects on Relaxational
Processes} Eds.: R. Richert and A. Blumen, (Springer, Berlin, 1994).
%
\bibitem{cummins94}
H. Z. Cummins, G. Li, W. M. Du and J. Hernandez, Physica A {\bf 204},
169 (1994).
%
\bibitem{yip95}
Theme Issue on Relaxation Kinetics in Supercooled Liquids-Mode Coupling
Theory and its Experimental Tests; Ed. S. Yip. Volume {\bf 24}, No.
6-8 (1995) of {\it Transport Theory and Statistical Physics}.
%
\bibitem{franosch96}
T. Franosch, W. G\"otze, M. Mayr and A. P. Singh, (unpublished).
%
\bibitem{colloids}
P. N. Pusey and W. van Megen, Phys. Rev. Lett. {\bf59}, 2083 (1987);
%
W. van Megen, S. M. Underwood and P. N. Pusey, Phys. Rev. Lett. {\bf
67}, 1586 (1991);
%
W. van Megen and S. M. Underwood, Phys. Rev. E {\bf 47}, 248 (1993);
%
W. van Megen and S. M. Underwood, Phys. Rev. Lett. {\bf 70}, 2766
(1993);
%
W. van Megen and S. M. Underwood, Phys. Rev. E {\bf 49}, 4206 (1994).
%
\bibitem{fluid_books}
J. P. Boon and S. Yip {\it Molecular Hydrodynamics} (Dover, New York, 
1980);
%
J.-P. Hansen and I. R. McDonald, {\it Theory of Simple Liquids}
(Academic, London, 1986).
%
\bibitem{goetze91}
W. G\"otze and L. Sj\"ogren, Phys. Rev. A {\bf 43}, 5442 (1991).
%
\bibitem{barrat90}
J.-L. Barrat and A. Latz, J. Phys.: Condens. Matter {\bf 2}, 4289
(1990).
%
\bibitem{bernu87}
B. Bernu, J.-P. Hansen, Y. Hiwatari and G. Pastore, Phys. Rev. A {\bf
36}, 4891 (1987).
%
\bibitem{roux89}
J. N. Roux, J.-L. Barrat and J.-P. Hansen, J. Phys.: Condens. Matter
{\bf 1}, 7171 (1989).
%
\bibitem{kob94}
W. Kob and H. C. Andersen, Phys. Rev. Lett. {\bf 73}, 1376 (1994).
%
\bibitem{kob95a}
W. Kob and H. C. Andersen, Phys. Rev. E {\bf 51}, 4626 (1995).
%
\bibitem{kob95b}
W. Kob and H. C. Andersen, Phys. Rev. E {\bf 52}, 4134 (1995).
%
\bibitem{kob95c}
W.~Kob and H.~C. Andersen, {\it Transport Theory and Stat. Physics},
{\bf 24}, 1179 (1995).
%
\bibitem{kob95d}
W. Kob and H. C. Andersen, Nuovo Cimento D {\bf 16}, 1291 (1994).
%
\bibitem{gotze85}
W. G\"otze, Z. Phys. B {\bf 60}, 195 (1985).
%
\bibitem{gotze87}
W. G\"otze, p. 34 in {\it Amorphous and Liquid Metals} Eds.: E.
L\"uscher, G. Frisch and G. Jacucci, (Martinus Nijhoff Publ.,
Dortrecht, 1987).
%
\bibitem{bosse87}
J. Bosse and J. S. Thakur, Phys. Rev. Lett. {\bf 59}, 998 (1987).
%
\bibitem{fuchs93}
M. Fuchs and A. Latz, Physica A {\bf 201}, 1 (1993).
%
\bibitem{fuchs_phd}
M. Fuchs, PhD Thesis, University of Munich, 1993.
%
\bibitem{fuchs_priv}
M. Fuchs, private communications.
%
\bibitem{bengtzelius86}
U. Bengtzelius, Phys. Rev. A {\bf 33}, 3433 (1986).
%
\bibitem{nauroth95}
M. Nauroth, Diploma Thesis (Mainz, 1995).
%
\bibitem{bgs}
U. Bengtzelius, W. G\"otze and A. Sj\"olander, J. Phys. C {\bf 17},
5915 (1984).
%



\end{references}
\end{document}